# Chapter 4

# Expanding World Views

## Can SETI expand its own horizons and that of Big History too?

**Michael A. Garrett**

> To understand the Universe, you must know about atoms - about the forces that bind them, the contours of space and time, the birth and death of stars, the dance of galaxies, the secrets of black holes. But that is not enough. These ideas cannot explain everything. They can explain the light of stars, but not the lights that shine from planet Earth. To understand these lights, you must know about life. About minds. (Stephen Hawking: speaking at the launch of the Breakthrough Listen Initiative).

**Abstract** The Search for Extraterrestrial Intelligence (SETI) is a research activity that started in the late 1950's, predating the arrival of "Big History" and "Astrobiology" by several decades. Many elements first developed as part of the original SETI narrative are now incorporated in both of these emergent fields. However, SETI still offers the widest possible perspective, since the topic naturally leads us to consider not only the future development of our own society but also the forward trajectories (and past histories) of many other intelligent extraterrestrial forms. In this paper, I present a provocative view of Big History, its rapid convergent focus on our own planet and society, its over-simplified and incomplete view of events in cosmic history, and its limited appreciation of how poorly we understand some aspects of the physical world. Astrophysicists are also not spared – in particular those who wish to understand the nature of the universe in "splendid isolation", only looking outwards and upwards. SETI can help re-expand all of our horizons but the discovery of extraterrestrial intelligence may also require its own practitioners to abandon preconceptions of what constitutes intelligent, sentient, thinking minds.

## 1 Introduction

For untold millennia, humankind has looked up at the sky and marvelled at the vastness and beauty of the cosmos. Countless generations have tried to understand their place in the centre of these immensities while contemplating the meaning of life and their own individual mortality. The scientific method has revealed the inner workings of the universe, and yet there are some fundamental questions that remain unanswered. One of these is: Are we alone?

Over the last few hundred years, astronomers have learned a great deal about the universe and our place in it. Major astronomical discoveries have been made, and almost without exception

Michael A. Garrett
Department of Physics & Astronomy
Jodrell Bank Centre for Astrophysics
University of Manchester, UK
e-mail: michael.garrett@manchester.ac.uk

these lend weight to the idea that life may be widespread in the universe. Most recently, Kunimoto and Matthews (2020) performed a statistical analysis of data collected by the Kepler space observatory, arguing that in our own Galaxy alone, there are more that 6 billion Earth-sized planets located in the so-called "habitable zone", orbiting Sun-like (G-type) stars. The evidence from our own planet's geological record demonstrates that simple life arose here very rapidly, evolving in ways that enabled life to survive and indeed prosper under quite diverse physical conditions. Exactly when life emerged on the Earth is a key input to Bayesian based analyses of this topic – if we accept depleted $^{13}$C in zircon mineral samples as the first evidence of life around 4.1 Gyr ago, then an objective analysis suggests that microbial life is likely to be common elsewhere in the Galaxy (Kipping, 2020). Whether it is reasonable to extrapolate from a sample of one is disputed – but historically, astronomers often do – the Copernican principle (also called the principle of mediocrity) is embraced rather strongly in our field, leading to a view in which life, and quite possibly intelligent life, is common.

We see a more cautious approach from other disciplines, in particular biology. By viewing the emergence of life as a random process, some biologists estimate the probability of life, and especially intelligent life as tiny, even compared to the billions upon billions of locations in the universe where the physical conditions might be right for it to arise (e.g. Cobb 2016). However, without a full understanding of how life actually originates (the murky transition between chemistry and biology), this also seems like a rather naïve standpoint. Probably the answer lies somewhere in between these two extremes. In the meantime, progress takes the form of two possible directions – either we sit and wait for biologists to generate a complete theory of life that culminates in a lab-based abiogenesis or we start searching for evidence for life outside of the Earth's biosphere now.

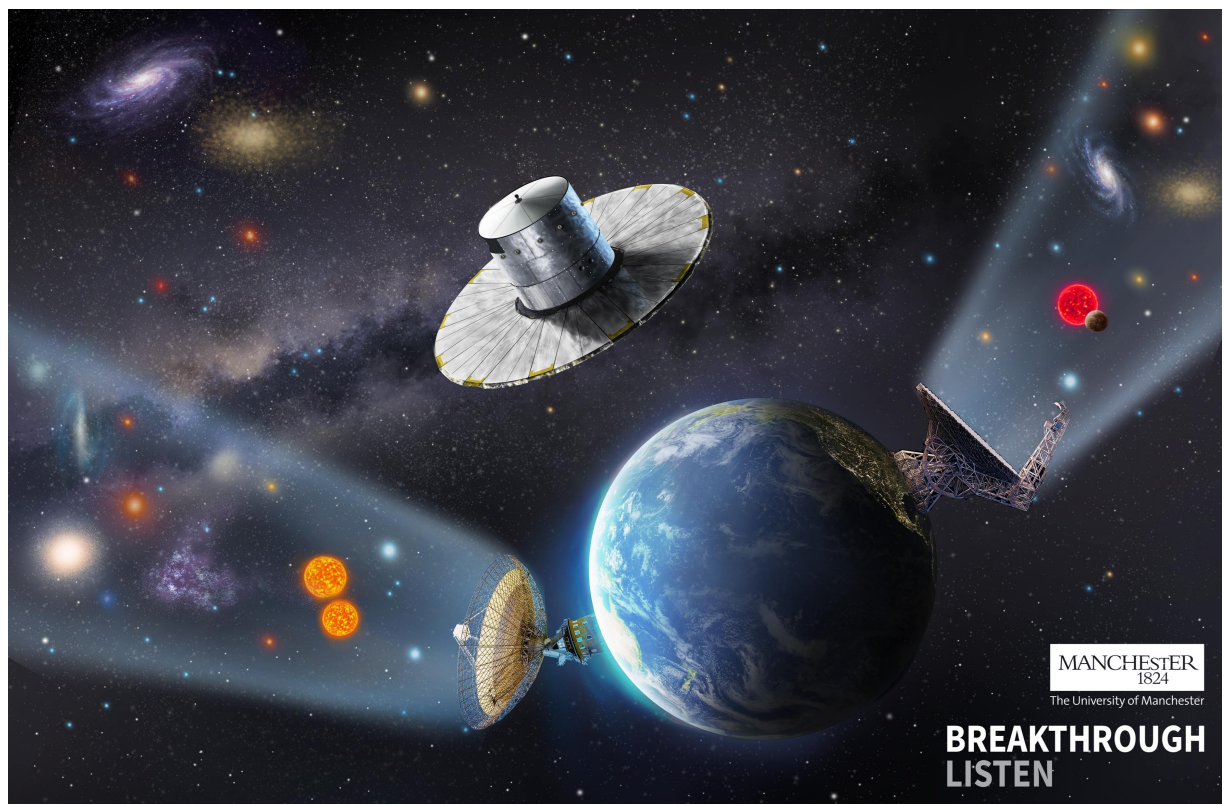

**Figure 1** Breakthrough Listen is placing the tightest limits yet on the prevalence of extraterrestrial civilisations, aided by data from the Gaia mission (Wlodarczyk-Sroka, Garrett and Siemion 2020). Artwork courtesy of Danielle Futselaar.

The Search for Extraterrestrial Intelligence (SETI) is a distributed research endeavour that carefully analyses astronomical data, looking for the tell-tale signatures of advanced technical civilisations located in our Galaxy and beyond (see Wright et al. 2019 for a recent review of the state of the profession). A wide variety of so-called "techno-signatures" are being searched for across the electro-magnetic spectrum, as recent advances in astronomical instrumentation have made us sensitive, for the first time, to anomalies in data associated with energy-intensive civilisations modifying their environments on planetary, stellar or galactic scales. Good examples include surveys that seek to detect waste heat (via excessive infrared emission) or extraterrestrial communication systems (via narrow-band radio signals). A renaissance in the field has recently taken place, partly inspired by the recent establishment of the privately funded Breakthrough Listen Initiative (Worden 2017). The first systematic SETI surveys for artificially generated, narrow-band radio signals are now underway, and a million-star survey using the MeerKAT radio telescope in South Africa is planned to commence next year. Recently, a re-analysis of the Breakthrough Listen radio data (Wlodarczyk-Sroka, Garrett and Siemion 2020) that also utilises stellar distances measured by the Gaia mission (see Fig.1), was able to place the tightest limits yet on the prevalence of high-duty-cycle extraterrestrial transmitters within 0.1-10 kpc (300-30,000 light years).

In parallel with these exciting developments, there is an enormous resurgence of interest in the topic of SETI world-wide. Young scientists are particularly attracted to the field, and the progress being made is also seen in the number and quality of SETI-related publications now appearing in well-established astronomical journals. With this dramatic blossoming of SETI related research, NASA itself has re-entered the field having recently organised a community workshop on the topic (Gelino et al. 2018).

## 2 Big History and SETI

The term "Big History" (hereafter BH) is attributed to Christian (1991) but many others have contributed to the topic. The idea is to attempt to describe the history of humankind within the context of a much bigger origins story that borrows heavily from the wide perspectives provided by cosmologists, astrophysicists, geologists, biologists, archaeologists, anthropologists etc. While there are many good textbooks on the topic (e.g. Spier 2010), the vast majority of students will first encounter the subject via online initiatives such as the Big History Project[1] (hereafter BHP). Funded by Bill Gates, BHP provides a stimulating course that freely provides high quality web-based resources to teachers and students alike. Themes in Big History include the importance of "Goldilocks conditions" (Spier 1996) and "thresholds of increasing complexity" (Christian 2008). These thresholds of increasing complexity are often used to sub-divide the topic into a number of easily digestible teaching modules (see Fig. 2).

### 2.1 Big History – a critique

The idea of providing a history course to high-school and university students that presents a much larger "Big Picture" view of history compared to traditional approaches is of course a good one. However, there are some aspects of the BHP courses that are troublesome in my view. In particular, the use of thresholds of complexity is rather artificial, and leads to some crucial aspects of cosmic evolution being rather neglected. In particular, the jump between threshold 3 (New Chemical Elements) and threshold 4 (Earth & The Solar System) represents a whopping 8+ Gyr of cosmic history! The result is that some very important and thematic phases in the evolution of the universe often get passed over – in particular:

- "Cosmic Dawn" - around 0.1-1 Gyr after the big bang at the end of the "dark ages" when the universe re-ionises with the emergence of large-scale structure in the form of the first stars, galaxies and proto galaxy clusters,
- "Cosmic Noon" – around 2-3 Gyr after the big bang where rapid galaxy evolution fuelled by mergers and gas accretion results in a significant peak in the cosmic star formation rate during which the vast majority of stars are formed and the metallicity of the universe begins to greatly increase. This period also sees the coeval growth of black holes which manifest themselves via spectacular Active Galactic Nuclei (AGN) that can outshine their galactic hosts, and
- The creation of our own Galaxy – with the bulge and halo forming first, followed by the disk in a process that was largely complete about 8 Gyr after the big bang.

---

[1] See https://www.bighistoryproject.com/home

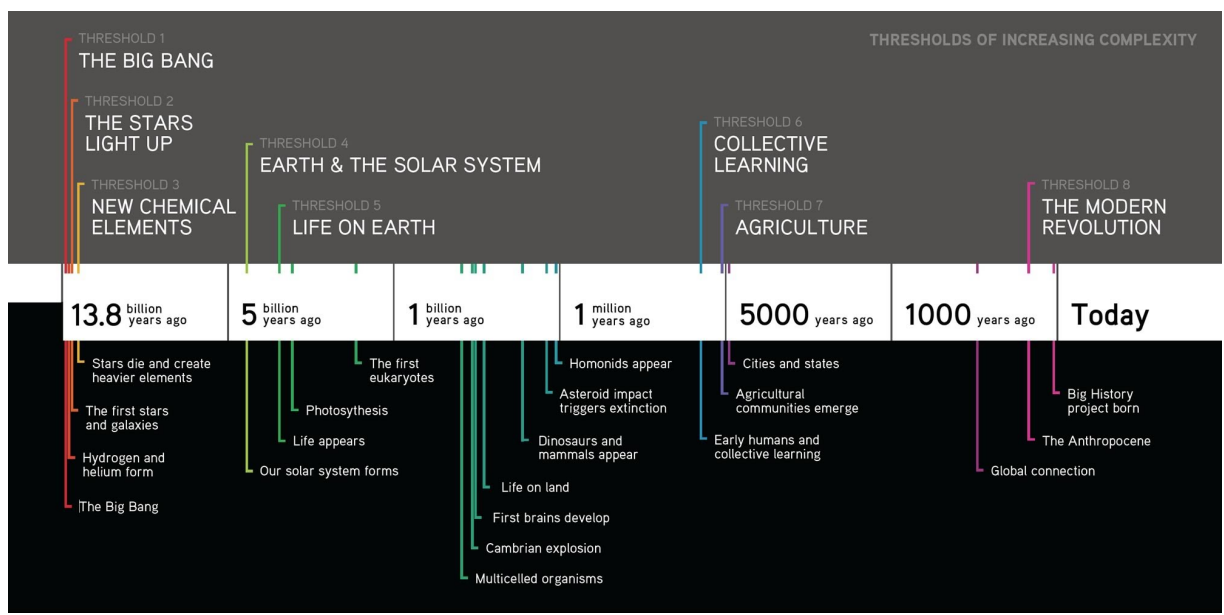

**Figure 2** The Big History Project proposes that history can be split up into thresholds of increasing complexity. (Image courtesy of the BHP teachers blog).

Having by-passed these important events, the focus of BH becomes entirely Earth-bound (see Fig. 2 – "Threshold 4") introducing the formation of our own solar system in the outskirts of the Milky Way about 5 Gyr ago. It therefore neglects to consider that countless other stellar systems have been forming planets (including rocky terrestrial ones) in our own and other galaxies for most of cosmic time (so for at least the last 10 billion years). Beyond this particular "threshold of complexity", the remaining BH narrative (see Fig. 2 thresholds 5-8) concentrates on the emergence and evolution of life on Earth, and the late emergence of human civilisation (sometimes considered to be commensurate with the first evidence for agrarian communities, around 10-20 thousand years ago). To my mind, the thresholds of increasing complexity that

BH identifies, seems to correlate not only with advancing cosmic time but with increasing introspection.

SETI takes a much broader view point, considering the possibility that life might be wide-spread throughout the universe, and that intelligent life, including advanced technical civilisations, may have developed not only on our own planet but on the countless others that have formed over the last 10 billion years of cosmic history. In that sense, SETI introduces a narrative that puts the "Big" back into an increasingly myopic "Big History". However, SETI also requires us to consider the lifetime of such technical civilisations – if they are all very short-lived (e.g. < 1000 years), the chances of them overlapping in both space and time becomes small (Drake 1961). For this reason, the longevity of a technical civilisation is a topic that is keenly considered by SETI researchers, with our own society serving as an example of what we might expect from others, or at least some others. In this sense, the future development of human civilisation becomes a key discussion point for SETI studies. Related to this, is the realisation that there is a lot of cosmic future still to come, even in our own Solar System that has at least another 5 Gyr of interesting evolution to look forward to – who knows what surprises there will be as the Sun begins to warm the outer regions of our own planetary system.

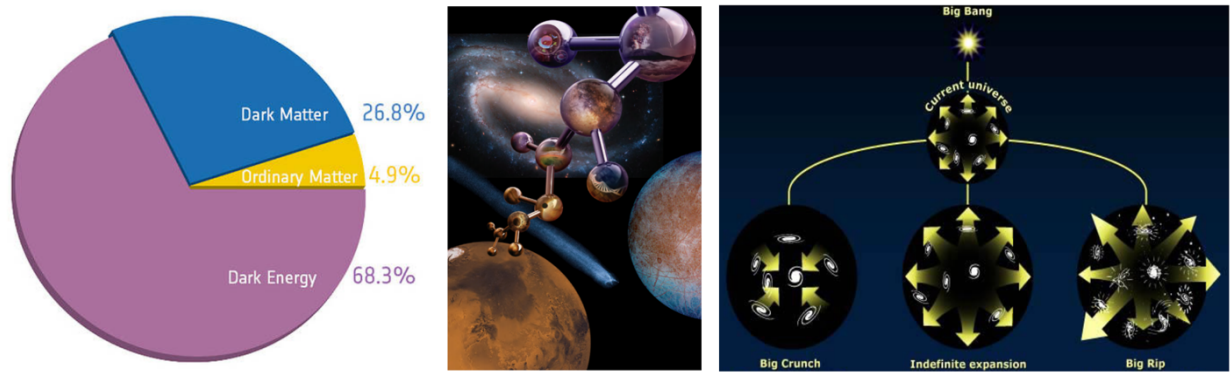

**Figure 3** A few examples of things we understand rather poorly – Left: The nature of dark energy and dark matter (presented here as "the pizza that no one ordered", image courtesy of ESA); Centre: Abiogenesis and the potential role of panspermia (image courtesy NASA); and Right: The ultimate fate of the universe (image courtesy of NASA).

While there is surely considerable benefit of exposing students to many of the topics covered in BH courses in terms of providing a well-rounded education, there remains the question of whether BH as a discipline itself provides any real added intellectual value. So far as I can see, the leading practitioners of BH have not made many unique contributions to our understanding of how human society will evolve over the coming years. It seems that domain specific studies by expert scientists bringing real depth and understanding to these issues do much better in this regard (e.g. Rees 2018).

Another shortcoming of BH is that the need to provide a very simplified (and, as we've seen earlier, incomplete) version of events, that can lead to a prescriptive narrative that ignores the fact that there are actually many aspects of the scientific world that we do not fully comprehend. A non-exhaustive list of some major topics for which our knowledge is sparse or totally lacking include:

- What is the nature of dark matter and dark energy – the primary components that dominate the energy budget of the universe, together with normal matter (see Fig. 3),
- Understanding inflation – the epoch just after the big bang in which the universe is required to experience a momentary phase of exponential expansion,
- What is the ultimate fate of the Universe – 'big freeze' *versus* 'big crunch' *versus* 'big rip'?
- How does life arise? (abiogenesis),
- Is there life beyond Earth (in the Solar System, other stellar systems etc)?
- Panspermia – can life propagate between planets in the same stellar system or even between the stars?
- Is there intelligent, conscious life beyond the Earth?
- Could we be living in a simulated multiverse in which physics and in particular quantum mechanics are the tools that generate levels of detail that currently equate to our own sense of reality and individual consciousness?

In the context of BH it is interesting to ask whether it matters that we do not know about these things in great detail. For example, is humankind's view of itself or of its future development greatly influenced by a deep knowledge of such things? The answer is almost certainly yes. For example, a steady-state theory of the universe would essentially guarantee that rare events (e.g. the emergence of intelligent life elsewhere) will always happen given sufficient time. On a more practical level, if we knew the answer to all, or even just a reasonable subset, of the questions listed above, we would probably be in a much better position to tackle some of the current challenges facing our own technical civilisation e.g. climate change, mass pandemics, energy production, general sustainability etc.

In any case, SETI can certainly help to widen BH's rather limited horizons. The idea that there may be other intelligent forms of consciousness 'out there' challenges us with a huge range of intriguing and curious questions. What will they look like? In what kind of environment have they evolved? Are they biological, machines or something else? Are they religious? Do they play football (soccer)? What kind of morals and ethics do they subscribe to? Do they create and enjoy music? What political ideologies do they subscribe to? Do they wage war against each other? Do they recognise the concept of love? What challenges did they overcome through the ages? How long do they live for? What do they eat? How do they reproduce? What has been their shared history? What can they teach us about mathematics, science, art and culture? There are so many questions to ask, and depending on the type of consciousness we encounter, some will be more relevant than others! What is quite clear is that any advanced technical civilisation will bring its own "Big History" to the table, and it promises to expand our minds in all sorts of different directions and unpredictable ways.

## *2.2 SETI and Astronomy*

Astrophysicists also have something to learn from SETI in my view – the very fact that intelligent life exists on this planet, and increasingly modifies its local environment, is something that should not be ignored as an irrelevant consideration. My experience is that there is still a substantial number of astronomers that wish to understand the nature of the universe in "splendid isolation" – only looking outwards and upwards, ignoring the fact that the presence of an intelligent, technical civilisation on our own planet (and potentially others) is surely telling us something rather profound about the nature of the universe we live in. This conservative approach sometimes leads to a blind application of physics that is almost designed from the outset to always exclude any non-natural explanation that might lead us to conclude that the imprint of energy-intensive civilisations might be found in astronomical data. A good example of how the community can over-react to a more expansive approach is the reception Bialy and Loeb (2018) received when they suggested that the non-Keplerian dynamics of the interstellar asteroid Oumuamua were consistent with it being a large artificial light sail (see Tingay 2020 for a discussion). My own view is that we should be encouraging such expansive thinking, provided it is backed up by good scientific evidence. In the words of the novelist Thomas Hardy, "While much is too strange to be believed, nothing is too strange to have happened" (see Taylor 1979).

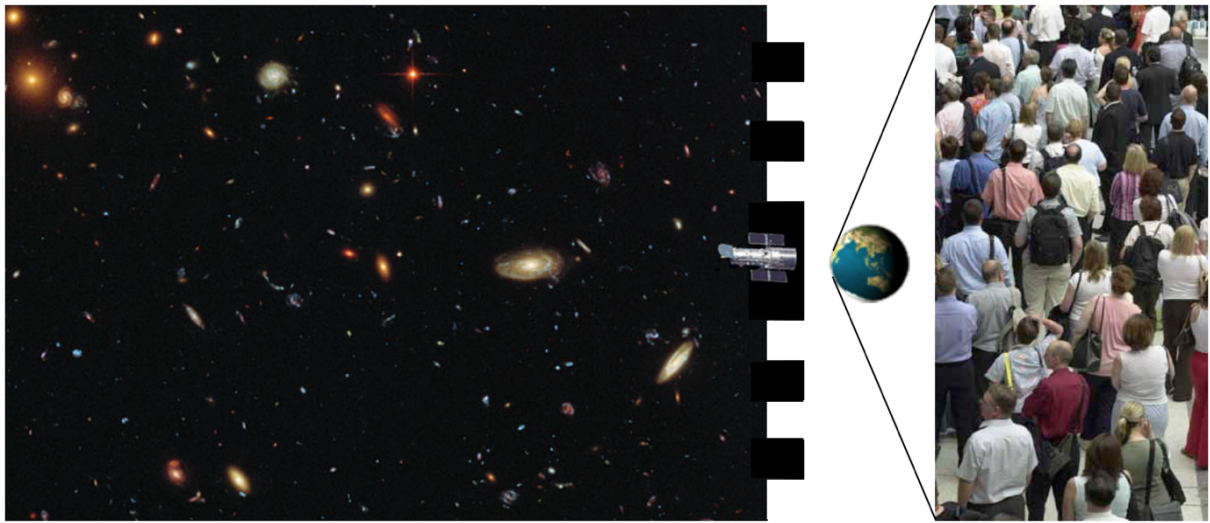

**Figure 4** Astronomers often view the universe in "splendid isolation"– only looking outwards and upwards – the interface with the living world around is often largely ignored (Image left is the Hubble Ultra Deep Field courtesy of NASA/ESA).

## 3 SETI success - thinking out of the box

This wider perspective that SETI demands, attracts contributions from many other disciplines. There are two reasons for this – first of all people from all walks of life are just extremely interested in the question of whether intelligent life exists elsewhere and what it will be like. And secondly, people coming from many other different areas of study (e.g. theology, music, linguistics, language and literature, medicine, economics, ethics, law, anthropology, AI, sociology, philosophy, technology, media studies, design, film, art, dance, sport, popular culture etc) understand that the discovery of another intelligent and thinking species could have a major impact on their own subjects.

As a result, SETI meetings often attract people from a wide range of different backgrounds, in addition to a hard core of scientists and technology developers. The variety of contributors is probably best represented by the SETI sessions organised annually at the International Astronautical Congress (IAC), in addition to meetings of the International Academy of

Astronautics (IAA) SETI Permanent Committee.[2] For those most engaged in the technical search, this variety of input can be a bit overwhelming at first but I personally find it highly stimulating, exercising our capacity to be open to all sorts of new ideas and critical thinking. The opinions of all of humankind are important to SETI – currently the scientific world cannot answer the question of whether we are alone – not even in our own Solar System - there is simply no consensus on this topic, so in my view, all opinions are valid, irrespective of where they originate from. In this sense, everyone has a legitimate and important contribution to make on this topic – at least for the moment. This is really important because one of the challenges facing SETI is our ability to recognise the presence of intelligence, especially if it comes in a form with which we are completely unfamiliar.

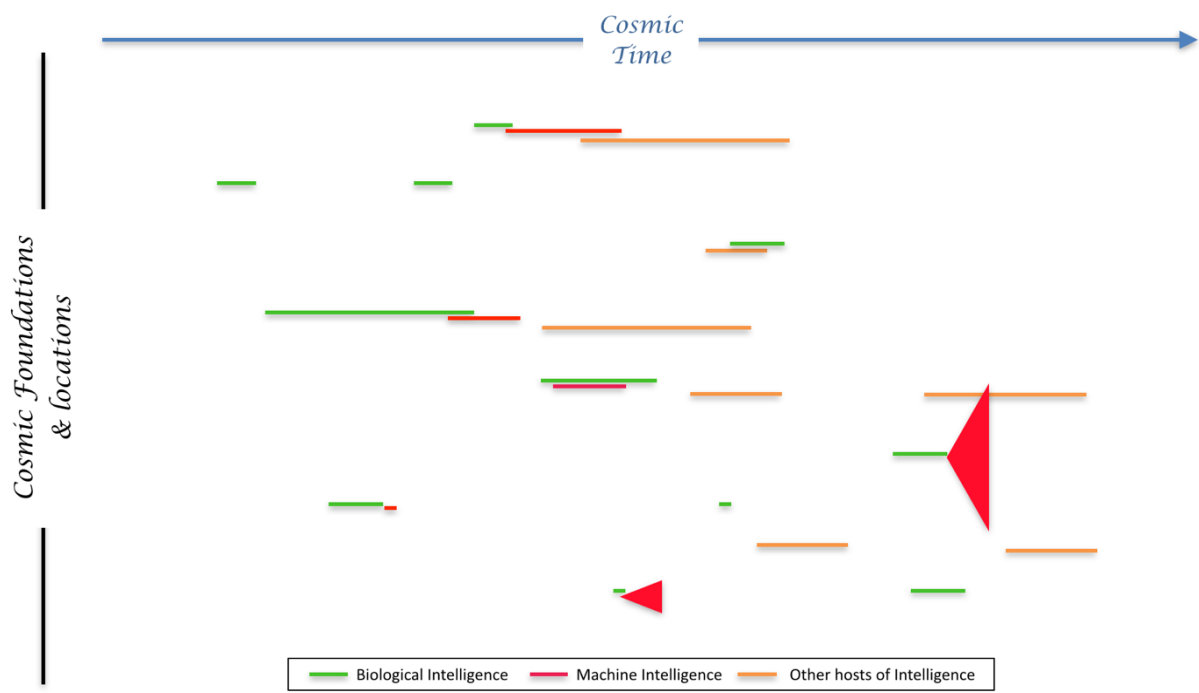

**Figure 5** A generic model in which intelligence arises at a range of times (horizontal axis) and location (vertical axis). Biological intelligence (green) may always be coupled with machine intelligence (red). Other potential hosts of intelligence that are uncoupled to biological intelligence are also presented (orange). Different forms of intelligence can overlap in time and location. Some forms of intelligence will expand out to occupy a range of locations (see triangular wedges). Occasionally some forms of intelligence (or their signatures) will overlap. A subset of possible examples is shown.

We already have been given some indications that the shape intelligence might take could be quite different from our preconceptions. When SETI searches began in the early 1960's, our focus was on the detection of technical civilisations very much like our own. At the same time, it was also appreciated that even on our own planet, our species did not have a monopoly on intelligence, e.g. dolphins and other marine mammals. In the last few decades, we have all observed the advances made in artificial intelligence (AI) and robotics, suggesting that intelligence and perhaps even consciousness itself might be embodied within non-biological entities (e.g. Kurzweil 1990; Bostrom 2015). These technologies, coupled with other advances such as nanotechnology, 3D-printing, advances in material sciences, molecular engineering, etc. can be very powerful in creating (for example) a microscopic intelligence that can reproduce itself easily and expand its presence well beyond our own solar system. As AI begins to play a major role in our own SETI searches (Zhang et al. 2018), the scenario of "our machines, detecting their machines" becomes increasingly more likely.

---
[2] https://iaaseti.org/en/

We can also reasonably ask whether there might be other non-biological ways of hosting intelligence, that we, as yet, have no knowledge of. We've only just begun to appreciate the role intelligent machines may play in our own future but there could be other non-biological hosts of intelligence that we (or they) have still to create or encounter. If Sagan (1980) was correct, and "we are a way for the cosmos to know itself", then it would be surprising if there were not other ways, other hosts of intelligence with the same ultimate driver. Some SETI scientists are uncomfortable with such broad thinking - the need to be "mainstream" in a field that already has its detractors (often other astronomers) can sometimes force us into an intellectual corner.

The main ideas are presented in a generic form in Fig. 5. Building on the enormous potential of our cosmic foundations (the big bang, the emergence of large-scale structure and stellar nuclear synthesis), there is an enormous potential for a wide range of different phenomena to arise, including conscious, intelligent forms. If our own example is typical, biological hosts of intelligence may be coupled with the rise of intelligent machines. Other non-biological hosts might exist independently of biological or machine intelligence. In the scenario depicted in Fig. 5, both biological and (unrelated) non-biological intelligence can arise at different times in the same location or (less likely) at the same time in the same location. Hosts of intelligence can also expand out from one location to settle many others (e.g. space faring societies). In these ways, intelligent hosts can have knowledge of each other, when they (or indeed their propagating signatures) overlap in space and time.

## 8 Conclusions

The wide perspective that is inherent in SETI research has much to offer the topic of BH. The idea that there may be many other forms of intelligence in our own Galaxy and beyond, encourages a much broader BH narrative that avoids an otherwise rapid descent towards a singular focus on our own planet and the development of one particular species on it. The idea of thresholds of increasing complexity seems to correlate not just with increasing cosmic time but also with increasing introspection. The incomplete nature of many online BH courses in terms of cosmic evolution should be addressed, in addition to the prescriptive way in which some aspects of the topic are taught. There are many aspects of the physical world that are completely opaque to us – we should not pretend otherwise. SETI opens our minds to consider some of these issues. At the very least we are intrigued about what shape these other forms of intelligence might take: will they reinforce our current thinking about ourselves and the universe, or will they bring a perspective that is totally disruptive, overturning some of the ideas and concepts we hold most firmly?

Astronomers can also benefit from a less blinkered view of the universe, and SETI researchers themselves must be prepared to think expansively, challenge our own preconceptions and open our mind to all possibilities, even if that means moving a bit further away from our own particular comfort zone. The idea that intelligence can also exist in non-biological hosts (including those that we have no familiarity or affinity with) is proposed as an example of the need to be prepared for the unexpected. In the words of the fictional character Fox Mulder, "the truth is out there".

**Acknowledgements** I'd like to thank Ian Crawford for organising a most enjoyable meeting at Birkbeck, University of London, on Expanding Worldviews, and also for helpful comments and suggestions on this text.